\documentstyle[prl,aps,multicol,epsfig,psfig,array]{revtex}

\begin{document}

\draft

\title{Bright soliton trains of trapped Bose-Einstein condensates}
\author{U. Al Khawaja,$^1$ H. T. C.  Stoof,$^1$
        R. G. Hulet,$^2$ K. E. Strecker,$^2$ and G. B. Partridge$^2$}
\address{
\it $^1$Institute for Theoretical Physics, Utrecht University,
Leuvenlaan 4, 3584 CE Utrecht, The Netherlands \\ $^2$Department
of Physics and Astronomy and Rice Quantum Institute, Rice
University, Houston, Texas 77251, USA}

\maketitle

\begin{abstract}
We variationally determine the dynamics of bright soliton trains
composed of harmonically trapped Bose-Einstein condensates with
attractive interatomic interactions. In particular, we obtain the
interaction potential between two solitons. We also discuss the
formation of soliton trains due to the quantum mechanical phase
fluctuations of a one-dimensional condensate.
\end{abstract}


\begin{multicols}{2}
{\it Introduction.} --- The existence of solitonic solutions is a
very general feature of nonlinear wave equations. Solitons,
therefore, play an important role in a large number of areas in
physics such as fluid dynamics, optics and high-energy physics. In
the case of an atomic Bose-Einstein condensate, the macroscopic
wave function of the condensate obeys the so-called
Gross-Pitaevskii equation, whose nonlinearity is a result of the
interatomic interactions. Depending on the repulsive or
attractive nature of the interatomic interactions, the
Gross-Pitaevskii equation allows for either {\it dark}, or {\it
bright} solitons, respectively. Dark solitons are associated with
a notch in the amplitude of the macroscopic wave function. Bright
solitons, on the other hand, correspond to localized wave
packets. The most important feature of both types of solitons is
that they do not disperse in time due to the nonlinearity in the
wave equation. The properties of dark solitons have been
extensively studied theoretically
\cite{zhang,reinhardt,dum,scott,jackson,muryshev,fedichev,james}.
They have also been created experimentally in elongated
Bose-Einstein condensates \cite{burger,denschlag,anderson}. Much
less is known of bright solitons \cite{carr}, which have only very
recently been created in two experiments with Bose-Einstein
condensates of $^7$Li atoms\cite{schreck,randy}.

In the experiment of Strecker {\it et al.}, soliton trains
consisting of up to 10 solitons have been observed \cite{randy}.
Moreover, it was found that even though the interatomic
interactions are attractive, the neighboring solitons repel each
other with a force that is dependent on their separation. Here we
confirm that the source of this repulsive force is a phase
difference of $\pi$ between two neighboring solitons
\cite{gordon}. Physically, this can be understood from the fact
that the antisymmetric nature of the many-soliton wave function
prevents the solitons from penetrating each other. Although the
phase difference explains the repulsive interactions between the
solitons, it presents us with the problem of understanding why
the experiment always seems to create soliton trains in which the
neighboring solitons have a $\pi$ phase difference. We also
address this interesting question here.

To understand the dynamics of the solitons we use a variational
approach, in which we describe the wave function of the individual
solitons as a gaussian and then make an appropriate linear
superposition of these gaussians to represent the soliton train.
Both the width and position of a gaussian can be varied. However,
since we are dealing with solitons, which by definition do not
show any dispersion, only the position is allowed to depend on
time. Using this trial wave function, we derive equations of
motion for the center-of-mass position and the relative distances
between the solitons. As expected, the center-of-mass motion
decouples completely from the relative motion. The dynamics of
the soliton separations compare favorably with the experimentally
observed behavior.

{\it Single soliton dynamics.} --- We first study 
the equilibrium properties and dynamics of a single
soliton. As in the experiments, we consider bright solitons
formed from a condensate that is confined by a very elongated
axially symmetric harmonic potential, for which the ratio of the
radial and the axial trapping frequency obeys
$\omega_{r}/\omega_z \gg 1$. We are, therefore, mostly interested
in the motion of the soliton in the axial direction and can
represent the wave function of a single soliton by a trial
function of the form
\begin{eqnarray}
\psi({\bf x},t) = A_{\rm sol}(z-\zeta(t),r)
                 \exp\left(i \frac{m}{\hbar} \frac{d\zeta(t)}{dt}
                 z\right)~,
\end{eqnarray}
where $\zeta(t)$ represents the soliton's center of mass, and $m$
is the atomic mass. For the amplitude of the soliton, we take the
gaussian
\begin{eqnarray}
A_{\rm sol}({\bf x})&=& \sqrt{\frac{N}{\pi^{3/2} q_z q_{r}^2}}
\exp\left(-{z^2\over2q_z^2}-{r^2\over2q_{r}^2}\right)
\label{trialsingle},
\end{eqnarray}
where $q_z$ and $q_r$ are two variational parameters that
determine the width of the soliton in the axial and radial
directions, respectively. The total number of atoms in the
condensate is denoted by $N$. Note that this gaussian {\it
ansatz} is particularly well-suited when the axial width of the
soliton is comparable to the axial harmonic oscillator length
$\ell_z=\sqrt{\hbar/m\omega_z}$. In the case that $q_z \ll
\ell_z$, it would be better to replace the exponential
$\exp(-z^2/2q_z^2)$ by $\sqrt{\pi}/[2\cosh(z/q_z)]$
\cite{gordon,schreck}. However, we will see later on that the
gaussian {\it ansatz} gives physically reasonable results in this
case also.

The equilibrium widths of the soliton can be calculated by
minimizing the Gross-Pitaevskii energy functional
\begin{eqnarray}
E[\psi^*,\psi]&=&\int d{\bf x}
\Big[{\hbar^2\over2m}|\mbox{\boldmath $\nabla$}\psi({\bf x},t) |^2
+V^{\rm ext}({\bf x})|\psi({\bf x},t)|^2\nonumber\\
&+&{1\over2}T^{\rm 2B}|\psi({\bf x},t)|^4\Big]
\label{energyfunctional},
\end{eqnarray}
with respect to $q_z$ and $q_r$ \cite{chris}. The external
potential obeys $V^{\rm ext}({\bf x}) = m(\omega_z^2 z^2 +
\omega_r^2 r^2)/2$ and the interatomic interaction strength
$T^{\rm 2B}=4\pi a\hbar^2/m$ is proportional to the negative
$s$-wave scattering length $a$. Assuming that $q_r$ has been
found in the above manner, the physics becomes effectively one
dimensional. The one-dimensional energy for $\psi(z,t) \equiv
\int dr 2\pi r \psi({\bf x},t)$ is again given by
Eq.~(\ref{energyfunctional}) but now we must use $T^{\rm 2B}=4\pi
\kappa\hbar^2/m$ with $\kappa=a/2\pi q_{r}^2$. It should be kept
in mind, however, that the one-dimensional theory can only be
applied for a soliton containing less than $N_{\rm max} = {\cal
O}(\ell_r/|a|)$ atoms, where $\ell_r = \sqrt{\hbar/m\omega_r}$ is
the radial harmonic oscillator length. Above this number of atoms
the soliton will collapse \cite{stoof,cass,ueda}.

Using this effective one-dimensional picture, the equation of
motion for $\zeta(t)$ is finally derived by substituting the
gaussian {\it ansatz} $\psi(z,t)$ into the action
$S[\psi^*,\psi]=\int dt (\int dz i\psi^*{\partial\psi/\partial
t}-E[\psi^*,\psi])$. The resulting equation $d^2\zeta(t)/dt^2 = -
\omega_z^2 \zeta(t)$ shows that the center of mass of the soliton
oscillates sinusoidially with the trap frequency. This is, of
course, an exact result that follows from the Kohn theorem
\cite{kohn}. 

{\it Two soliton dynamics.} --- Our trial wave functions for two
solitons with a phase difference of $0$ or $\pi$ are,
respectively, the symmetric and antisymmetric combinations of the
two gaussian wave functions of a single soliton. In detail we use
\begin{eqnarray}
&&\psi_{\pm}(z,t)= \nonumber \\ &&\hspace{0.1in}
\frac{1}{\sqrt{N_{\pm}(t)}}\Bigg[ A_{\rm sol}(z-z_1(t))
    \exp\left(i \frac{m}{\hbar} \frac{dz_1(t)}{dt}
                (z-\zeta(t))\right)\nonumber\\
&&\hspace{0.4in} \pm A_{\rm sol}(z-z_2(t)) \exp{\left(i
\frac{m}{\hbar} \frac{dz_2(t)}{dt} (z-\zeta(t)) \right)}\Bigg]
\label{trial}.
\end{eqnarray}
Here $z_1(t)$ and $z_2(t)$ denote the positions of the two
solitons and $\zeta(t)=[z_1(t)+z_2(2)]/2$ is their center of mass.
For simplicity we consider here only the case of a phase
difference of $0$ or $\pi$ between the solitons. In addition, we
take the number of atoms in each soliton, and therefore also their
widths, equal to each other. The generalization to an arbitrary
phase and an arbitrary number of atoms is straightforward but
somewhat tedious. Fortunately, the above assumptions turn out to
be reasonable when we compare our results with experiments, and it
allows us to bring out the dynamics of the relative separation
most clearly. Finally, for the rest of this section we scale
length to $\ell_z=\sqrt{\hbar/m\omega_z}$, time to $1/\omega_z$,
and energy to $\hbar\omega_z$. The normalization constant
$N_{\pm}(t)$ is then given by
\begin{equation}
N_{\pm}(t)=2\left(1\pm
\exp{\left[-{\eta^2(t)\over4q_z^2}-{q_z^2\over 4}
\left(\frac{d\eta(t)}{dt}\right)^2 \right]}\right) \label{Apm},
\end{equation}
where $\eta(t)=z_1(t)-z_2(t)$ is the distance between the two
solitons. Note that $N$ is still the total number of atoms in the
condensate, so a single soliton contains only $N/2$ atoms.

Using this trial wave function, the lagrangian per atom takes the
form
\begin{eqnarray}
\frac{L[\zeta,\eta]}{N}&=&{1\over2}\left[\left(\frac{d\zeta}{dt}\right)^2-
\zeta^2\right]+{1\over8}\left[\left(\frac{d\eta}{dt}\right)^2-
\eta^2\right]\nonumber\\ &-&{1\over4}\left[{1\over q_z^2}+q_z^2 +
V[\eta] \right] \label{lag1},
\end{eqnarray}
where the effective potential $V[\eta]$ is given by
\begin{eqnarray}
&&\hspace{-0.1in}V[\eta]= \nonumber\\
&&\mp{1\over
N_\pm}\left[(1+q_{z}^{-4})\eta^2+(q_{z}^4-3){\dot\eta}^2\right]
\exp{\left(-{\eta^2+q_{z}^4{\dot\eta}^2\over4q_{z}^2}\right)}
\nonumber\\
&&+{8\sqrt{2\pi}N\kappa\over N_\pm^2q_{z}} \Bigg[1 +
\exp{\left(-{\eta^2+q_{z}^4{\dot\eta}^2\over2q_{z}^2}\right)}
+2\exp{\left(-{\eta^2\over2q_{z}^2}\right)}\nonumber\\ &&\pm
4\exp{\left(-{3\eta^2+q_{z}^4{\dot\eta}^2\over8q_{z}^2}\right)}
\cos{\left({\eta{\dot\eta}\over4}\right)}\Bigg]
\label{effectivepotential},
\end{eqnarray}
and $\dot\eta = d\eta/dt$. It is clear from this lagrangian that
the center-of-mass motion and the relative motion are completely
decoupled. Furthermore, the equation of motion for the center of
mass is simply $d^2\zeta(t)/dt^2 = - \zeta(t)$, which corresponds
again to the Kohn mode that oscillates at the frequency of the
trap. On the other hand, the equation of motion for the relative
motion is rather complex due to the coupling between the two
solitons that is determined by the velocity-dependent potential
$V[\eta]$.

\begin{figure}
\epsfig{figure=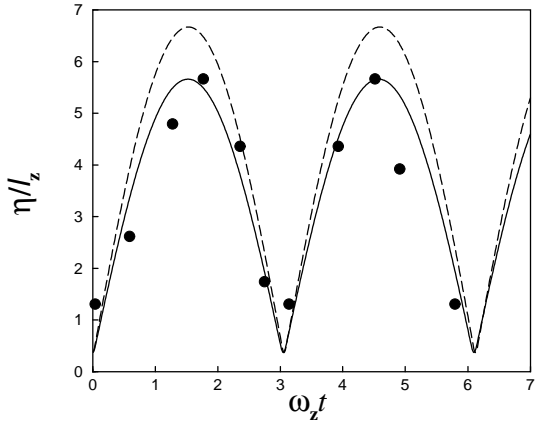,width=7.0cm}
\caption{\narrowtext The relative motion of two solitons. There
are 5000 atoms in each soliton and $a=-3a_0$, with $a_0$ the Bohr
radius. The solid curve is the solution of the exact equation of
motion for $\eta(t)$. The dashed curve is the solution of
Eq.~(\ref{etam}) with $V(\eta)=(\eta^2/2q_z^4)\exp(-\eta^2/4
q_z^2)$. The points are experimental data obtained using the same
apparatus described in Ref.~[14]. In the experiment more than two
solitons were created. We focused on two large and adjacent
ones.} \label{fig1}
\end{figure}

In Fig.~\ref{fig1}, we compare the exact numerical solution of
the relative motion with experimental data. The parameters of the
experiment and the theory are the same as those of
Ref.~\cite{randy}, and we take the anti-symmetric combination of
gaussians. Very good agreement is obtained with both the
experimentally observed amplitude of oscillation of the relative
coordinate and its frequency, which is approximately $2\omega_z$.
In particular, we see that the two solitons never cross each
other. This result, and the nonsinusoidal nature of $\eta(t)$,
are both due to the hard-core nature of the interaction when the
solitons start to overlap.

To show this, we neglect the velocity dependence of $V[\eta]$,
assume that $q_z \ll 1$ and take the limit of large distances
between the two solitons, i.e., $\eta \gg q_z$. With these
approximations we obtain
\begin{equation}
V[\eta]={2\sqrt{2\pi}N\kappa \over q_{z}} \mp
\left({4\sqrt{2\pi}N\kappa \over q_{z}} + {\eta^2\over2q_z^4}
\right) \exp{\left(-{\eta^2\over4q_z^2}\right)} \label{effapp}.
\end{equation}
The second term in the right-hand side represents the interaction
energy $V(\eta)$ between the two solitons. For the antisymmetric
choice of the two-soliton wave function this potential energy is
positive, which implies that the two solitons will indeed repel each
other.

In this approximation, the equation of motion of $\eta(t)$ takes
the form of Newton's equation
\begin{equation}
\frac{d^2 \eta(t)}{dt^2}
  =-\eta(t) - \frac{d V(\eta(t))}{d \eta}
\label{etam}\;.
\end{equation}
Thus, the problem reduces to that of a particle moving in the
potential $\eta^2/2+V(\eta)$, which has two minima located
symmetrically around $\eta=0$. Depending on the initial
conditions, the two solitons can be trapped in these minima. In
other words, the separation between them oscillates and the
solitons never cross each other, as verified by the exact
solution. To investigate the importance of the interatomic
interactions, we show in Fig.~\ref{fig1} with the dashed curve,
the solution to Eq.~(\ref{etam}) when the potential contains only
the contribution from the kinetic energy of the atoms. A solution
of Eq.~(\ref{etam}) with the full potential $V(\eta)$ from
Eq.~(\ref{effapp}) is almost identical to the exact solution. The
smaller amplitude of oscillation in the relative motion is,
therefore, a clear signature of the reduction of the repulsive
force between the solitons due to the interatomic interactions.
We have also analytically calculated the
interaction energy for two solitons with an amplitude
proportional to $\sqrt{\pi}/[2\cosh(z/q_z)]$ and solved the
corresponding equation of motion. It leads to only minor
corrections that are due to the fact that the potential $V(\eta)$
is now somewhat more repulsive and only falls off as
$(4\eta/q_z^3)\exp(-\eta/q_z)$ for large separations.

{\it Soliton train formation} --- In the experiment of Strecker
{\it et al.} \cite{randy}, soliton trains are formed by making use
of a so-called Feshbach resonance \cite{eite}. In summary, the
procedure consists of first making a stable condensate with a
relatively large positive scattering length, and then switching
to a small negative value. An important clue for understanding
the mechanism for the creation of the soliton train is the
observation that all the solitons repel each other, and thus have
a phase difference equal to $\pi$. This suggests that by changing
the sign of the scattering length, the phase modes of the
condensate become unstable. Roughly speaking, the most important
wavelength of the unstable phase modes must then be comparable to
the separation between solitons in the train.

To make this physical idea quantitative, we consider a
homogeneous condensate with a density $n=N/L$, where $L$ is the
(Thomas-Fermi) length of the condensate. Denoting the phase of the
condensate wave function $\psi(z,t)$ by $\chi(z,t)$, it is easy to
show from the action $S[\psi^*,\psi]$ that $\chi_k(t) \equiv \int
dz~e^{-ikz} \chi(z,t)$ corresponds to a (complex) harmonic
oscillator for every wave vector $k>0$. Physically this implies
that the quantum mechanics of every phase mode is analogous to
the quantum mechanics of a fictitious particle in a
two-dimensional harmonic oscillator potential. The mass of this
fictitious particle is $m_k = 4\epsilon_kn/\omega_k^2L$ and the
frequency of the harmonic oscillator potential is $\omega_k =
\sqrt{\epsilon_k^2 + 2nT^{2B} \epsilon_k}/\hbar$, with
$\epsilon_k=\hbar^2k^2/2m$. Initially the scattering length is
positive and all these frequencies are positive. Assuming that
the condensate has come to equilibrium, the quantum fluctuations
of $\chi_k(t)$ are then fully determined at zero temperature by
the ground-state wave function $(1/\pi \ell_k^2) \exp(-
|\chi_k|^2/2\ell_k^2)$, with $\ell_k = \sqrt{\hbar/m_k\omega_k}$.
In particular, we have that $\langle \chi(z,t) \rangle = 0.$

However, if at $t=0$ we suddenly change the scattering length to a
negative value, the modes with $k<k_{\rm max}=\sqrt{16\pi|\kappa|
n}$ become unstable because the spring constant $m_k \omega_k^2$
of these harmonic oscillators becomes negative and equal to $- m_k
\Omega_k^2$. Solving the quantum mechanics of a particle in an
inverted harmonic oscillator potential, we can then show that for
$t \gg 1/\Omega_k$
\begin{equation}
\langle |\chi_k(t)|^2 \rangle \simeq \frac{\hbar}{4 m_k \omega_k}
   \left( 1 + \frac{\omega_k^2}{\Omega_k^2} \right) e^{2\Omega_k
   t}~.
\end{equation}
As in the case of a spontaneously broken symmetry, the latter
result can physically be understood by saying that quantum
mechanical fluctuations {\it imprint} the condensate wave
function with the phase
\begin{equation}
\langle \chi(z,t) \rangle \simeq
  \int_0^{k_{\rm max}} \frac{dk}{2\pi}\; \cos(kz)
  \sqrt{\frac{\hbar\omega_k L}{4\epsilon_k n}
        \left(1 + \frac{\omega_k^2}{\Omega_k^2} \right)}
        e^{\Omega_k t}~.
\label{imprint}
\end{equation}

\begin{figure}
\begin{center}
\epsfig{figure=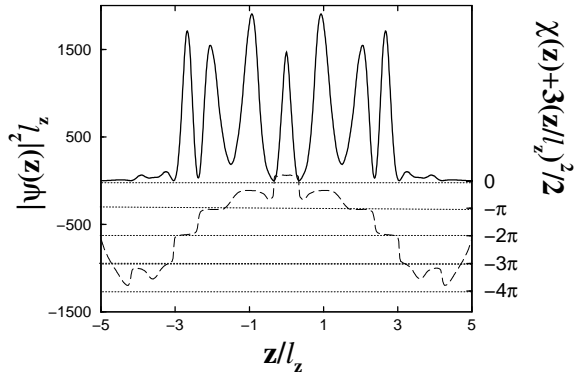,width=7.5cm}
\end{center}
\caption{\narrowtext Soliton train formation. The solid curve is
the density and the dashed curve is the phase of the condensate.
Initially, we start with an equilibrium condensate profile for
$10^4$ atoms with a scattering length of $200 a_0$. The
scattering length is then instantaneously changed to $a=-3 a_0$.
The trap parameters are obtained from Strecker {\it et al.} [14].}
\label{fig2}
\end{figure}

Having obtained this result, we are now able to simulate the
experiments. We first determine the ground state wave function of
a condensate with a positive scattering length for the
experimental parameters of interest. We then imprint this wave
function with the phase given in Eq.~(\ref{imprint}) and let the
resulting wave function evolve under the appropriate
Gross-Pitaevskii equation with a negative scattering length. We
note here that, due to numerical limitations, we have used an
initial number of $10^4$ atoms which is an order of magnitude less
than that in the experiment. In this case, we find that after an
evolution time of $t \simeq 1.8/\omega_z$, seven solitons are
formed, together with phonon excitations. 
The solitons are clearly visible as plateaus in the phase, after subtraction 
of the parabolic behaviour of the phase due to the overall shrinking of the condensate. 
The phase difference between two
adjacent solitons is indeed approximately equal to $\pi$. 
Some deviations from $\pi$ are expected because the solitons are moving 
relative to each other. These
results are shown in Fig.~\ref{fig2}, where we plot both the
density and the phase of the condensate wave function. Note that
our calculation includes no relaxation processes due to the
presence of a thermal cloud \cite{duine}. Including these
stabilizes the soliton train by damping the phonon excitations 
and the soliton motion.

{\it Conclusions} --- In this Letter, we have considered the
formation and dynamics of bright soliton trains in a trapped
Bose-Einstein condensate with attractive interactions. In
particular, we have shown that two solitons with a phase
difference of $\pi$ repel each other. The resulting dynamics
compares favorably with experiment. In the experiment, always
more than two solitons were created, while here we have focused on
only two. In principle, a generalization to more than two
solitons is straightforward. In this case, we just sum in the
lagrangian of Eq.~(\ref{lag1}) over all the positions $z_i(t)$ of
the bright solitons.

We have also shown that soliton trains can be dynamically
generated by quantum mechanical phase fluctuations of the
condensate, due to the modulational instability \cite{tai} that
exists in a Bose-Einstein condensate with attractive interactions.
This mechanism gives an intuitive explanation why adjacent
solitons have a $\pi$ phase difference. It also predicts that the
number of solitons that is formed after a rapid, i.e., fast
compared to $1/\omega_z$, sign change in the scattering length, is
equal to the ratio of the initial condensate size and the
wavelength of the most unstable phase mode. However, this
prediction does not take dissipation into account. Because
solitons are not topological objects, dissipation can lead to a
considerably different number of solitons as it drives the gas to
equilibrium. In the experiment, the number of solitons in the
train is about half of the dissipationless prediction
\cite{randy}. This observation suggests that dissipation
processes are indeed present. A detailed comparison with
experiment thus requires knowledge of the properties of the
thermal cloud, which are not well known experimentally. It was
also observed that the number of solitons in the train is, over a
large range, independent of the exponential time constant with
which the scattering length is changed. This is as expected as
long as dissipation plays an important role and the sign change in
the scattering length remains nonadiabatic.

\end{multicols}

\end{document}